\documentclass[pra,reprint,amssymb]{revtex4-1}%

\usepackage{amsmath,amssymb}
\usepackage{revsymb4-1}
\usepackage[cmyk]{xcolor}
\usepackage[%
	colorlinks=true,%
	linkcolor=blue,%
	citecolor=blue,%
	urlcolor=blue
	]{hyperref}%
\usepackage{graphicx}%
\usepackage{here}
\usepackage{dcolumn}%
\usepackage{bm}%
\usepackage{revsymb}
\usepackage{braket}
\usepackage{layouts} 
\usepackage{color}
 
\def\equationautorefname~#1\null{\textrm{~(#1)\;}\null}
\def\figureautorefname~#1\null{Fig.~#1\null}

\begin{document}

\title{Hidden multiparticle excitation \\ in weakly interacting Bose--Einstein Condensate}
\author{Shohei Watabe}

\begin{abstract}
We investigate multiparticle excitation effect on a collective density excitation as well as a single-particle excitation in a weakly interacting Bose--Einstein condensate (BEC). We find that although the weakly interacting BEC offers weak multiparticle excitation spectrum at low temperatures, this multiparticle excitation effect may not remain hidden, but emerges as bimodality in the density response function through the single-particle excitation. Identification of spectra in the BEC between the single-particle excitation and the density excitation is also assessed at nonzero temperatures, which has been known to be unique nature in the BEC at absolute zero temperature. 
\end{abstract}

\affiliation{Tokyo University of Science, 1-3 Kagurazaka, Shinjuku-ku, Tokyo, 162-9601, Japan}

\pacs{03.75.Kk, 67.85.De}
\maketitle

Bose--Einstein condensates (BECs) --- non-vanishing nature of off-diagonal long-range order for the one-body density matrix~\cite{Penrose1956,Yang1962} --- have drawn considerable research interest in many branches of physics ranging from equilibrium states, such as liquid helium~\cite{London1938,Kapitza1938,Allen1938,London1938}, ultracold atomic gases~\cite{Anderson1995,Davis1995} as well as magnetic insulators~\cite{Nikuni2000,Ruegg2003,Giamarchi2008}, to nonequilibrium states, such as magnon~\cite{Demokritov2006}, phonon~\cite{Misochko2004}, photon~\cite{Klaers2010}, polariton~\cite{Balili2007} as well as exciton-polariton~\cite{Kasprzak2006,Plumhof2014}. In superfluid helium, in particular, multiparticle excitations have been extensively studied in the context of interacting maxons as well as rotons~\cite{Talbot1988,Glyde1990,Glyde1992PRB,Ruvalds1970,Zawadowski1972,Griffin1973,Manousakis1986,Fukushima1988,Juge1994}, since their bound states as well as scattering states are quite important because of their very large density of states~\cite{Ruvalds1970,Zawadowski1972,Griffin1973,Manousakis1986,Fukushima1988,Juge1994}. 

In recent BEC systems, on the other hand, such as dilute ultracold monoatomic BECs~\cite{Anderson1995,Davis1995} as well as quasi-particle BECs~\cite{Nikuni2000,Ruegg2003,Giamarchi2008,Kasprzak2006,Plumhof2014}, spectra measured in those experiments show the Bogoliubov-type phonon dispersion relation in the low-momentum regime~\cite{Stamper-Kurn1999,Steinhauer2002,Ruegg2003,Utsunomiya2008}, not showing maxon and roton branches. If we employ our understanding of multiparticle excitations in liquid helium, multiparticle excitations would not be important in those BECs, because these do not manifest quasi-particle branches enhancing the density of states, where the dispersion relation becomes flat~\cite{Ruvalds1970,Zawadowski1972,Griffin1973,Manousakis1986,Fukushima1988,Juge1994}. However, owing to the presence of the BEC, a single quasi-particle can create two quasi-particles through an interaction process with the BEC~\cite{griffin1993excitations}, which might provide multiparticle effect. In the end, we are confronted with a vital question about the multiparticle excitation in BECs, which have recently drawn considerable research interest, without maxons as well as rotons. 

In this paper, we tackle this problem, and unveil multiparticle excitation effect in a weakly interacting Bose gas. We show that in the presence of the Bogoliubov-type excitation even without maxons and rotons, weak multiparticle excitation effect is no longer hidden, but emerges as a bimodal structure in the density response function through the single-particle excitation. This is evidence of the many-body effect in the BEC; since the quasi-particle excitation and the density excitation are hybridized thanks to the BEC~\cite{Gavoret1964,griffin1993excitations}, a weak multiparticle excitation effect is included into the single-particle excitation through the interaction between the condensate and the quasi-particle in the density-fluctuated medium affected by the weak multiparticle excitation. For bimodality, quasi-particle--quasi-particle interaction effect in the many-body medium is not found to be important, because the many-body medium effect is smeared out by quasi-particles with various momenta. We also discuss the relation of peaks in the single-particle spectral function and the density response function. Although at extremely low temperatures, these peaks are merged, which is consistent with the Gavoret-Nozi\`eres prediction~\cite{Gavoret1964}, the peak lines are distinguishable at higher temperatures, since less is the weight of those density vertices which involve the single-particle Green's function into the density response function. 

Consider a weakly interacting Bose gas with the Hamiltonian 
\begin{equation}
    H=\sum\limits_{\mathbf{p}}  (\epsilon_\mathbf{p} - \mu)a_\mathbf{p}^\dag a_\mathbf{p} + {U\over2} \sum\limits_{\mathbf{p},\mathbf{p}',\mathbf{q}} a_{\mathbf{p}+\mathbf{q}}^\dag a_{\mathbf{p}'-\mathbf{q}}^\dag a_{\mathbf{p}'} a_\mathbf{p} ,  \label{eq-H}
\end{equation}%
where $a_\mathbf{p}$ is the annihilation operator of a bosonic particle with a mass $m$ as well as the kinetic energy $\epsilon_\mathbf{p} = {\bf p}^2/(2m)$ measured from the chemical potential $\mu$. A weak repulsive interaction $U (>0)$ is described by an $s$-wave scattering length $a$ such that $4 \pi a/m = U/[1+U\sum_\mathbf{p} (2 \epsilon_\mathbf{p})^{-1}]$. The BEC order parameter $\langle a_{\mathbf{p} = {\bf 0}} \rangle = \sqrt{n_0} $ is assumed to be real without loss of generality, where $n_0$ is the condensate density.  (In this paper, we use the convention such that $\hbar$, $k_{\rm B}$ and a volume are unity.)

The self-energy $\Sigma(p)$ includes interaction effect in density-fluctuated medium. (We have simply given $p \equiv (i \omega_n, {\bf p})$, where $\omega_n$ is the boson Matsubara frequency.)  Below the critical temperature, this density fluctuation is formed by the quasi-particles with the phonon dispersion relation in the low-energy regime. In order to include this effect, we employ, for constructing the self-energy $\Sigma (p)$ as well as the other building blocks, the Green's function in the Hatree-Fock-Bogoliubov-Popov approximation~\cite{Popov1964A,Griffin1996}, which successfully describes the Bogoliubov phonon dispersion relation in the low-energy regime. (Details of our theory used in this paper are summarized in the Supplemental Material.)

The self-energy contribution in the BEC phase is constructed from condensate parts $\Sigma_{11(12), {\rm c}}$ as well as non-condensate parts $\Sigma_{11(12),{\rm n}}$. In the case of  the Hatree-Fock-Bogoliubov-Popov-type diagrammatic contribution~\cite{Popov1964A,Popov1983,Griffin1996,Watabe2013}, one can employ $\Sigma_{11} (p) = \Sigma_{11, \rm c} (p) + \Sigma_{11,\rm n} (p)$ as well as $\Sigma_{12} (p) = \Sigma_{12,\rm c} (p)$. 
The density response function $\chi(p)$ below the BEC critical temperature $T_{\rm c}$ is constructed by the two parts $\chi (p) = \chi^{\rm 1PI}(p) + \chi^{\rm 1PR} (p)$~\cite{griffin1993excitations,Gavoret1964}, where $\chi^{\rm 1PI} (p)$ is the one-particle irreducible (1PI) part, and $\chi^{\rm 1PR} (p) = \Upsilon^{\dag} (p) G(p) \Upsilon(p)$ is the one-particle reducible (1PR) part that is specific to the BEC phase. Here, $\Upsilon$ and $\Upsilon^{\dag}$ are density vertices that involve into the density response function the single-particle Green's function $G^{-1} (p) = i \omega_n \sigma_3 - \epsilon_{\bf p} + \mu - \Sigma(p)$. (Details are summarized in the appendix.)

\begin{figure}[tb] 
	\centering
\includegraphics[width=7cm]{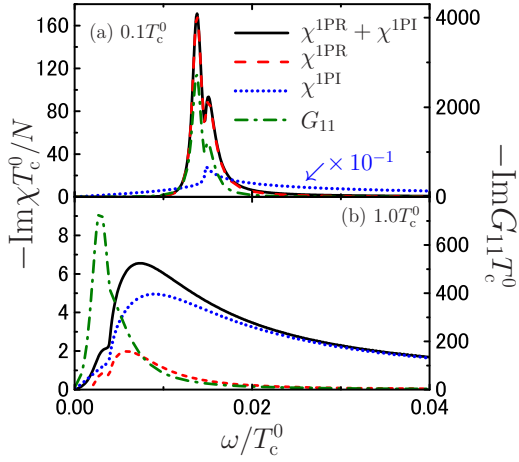}
\caption{
Density response function and single-particle spectral function at $ T = 0.1T_{\rm c}^{0}$ (a) and at $T_{\rm c}^{0}$ (b). Here, $T_{\rm c}^{0}$ is the BEC critical temperature of an ideal Bose gas. We have used the gas parameter $a n^{1/3} = 10^{-2}$, where the critical temperature in this system is given by $T_{\rm c} /T_{\rm c}^{0} \simeq 1 + 1.9 an^{1/3}$. All the spectral functions are at $q = 0.05 q_{0}$, where $T_{\rm c}^{0} \equiv q_{0}^{2}/(2m)$. 
 }
\label{fig1}
\end{figure}

The density response function shows the crossover from the (BEC-specific) 1PR-dominant regime (Fig.~\ref{fig1} (a)) to the (BEC-nonspecific) 1PI-dominant regime (Fig.~\ref{fig1}(b)). In the low-temperature regime, the density response function exhibits a sharp peak, which originates from the 1PR part, {\it i.e.,} the single-particle Green's function. This single-particle contribution to the density response function is due to the presence of the BEC providing the nonzero density vertices $\Upsilon$ and $\Upsilon^{\dag}$.  Compared with the 1PR part, the 1PI part is negligibly small, where a sharp point is positioned close to the single-particle excitation peak. At higher temperatures, on the other hand, since density vertices possess the factor $\sqrt{n_{0}}$ through the condensate Green's function, a strong sharp peak disappears in the 1PR part. The 1PI part thus becomes relatively dominant at higher temperatures, although the sharpness in $\chi^{\rm 1PI}$, as seen in lower-temperatures, disappears. The contribution of $\chi^{\rm 1PR}$ to the whole $\chi$ completely vanishes at $T = T_{\rm c}$, because of the absence of the hybridization.

\begin{figure}[tb] 
\centering
\includegraphics[width=3.5in]{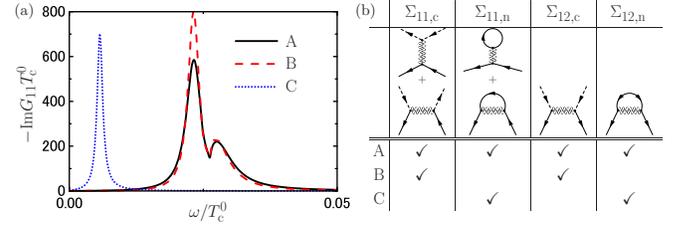}
\caption{
Single-particle spectral function with various approximations. (a) The single-particle spectral function at $T = 0.5 T_{\rm c}^{0}$ with $q = 0.1 q_{0}$, where $an^{1/3} = 10^{-2}$ is used. (b) Diagrammatic contributions used in (a). The solid arrow represents the single-particle line, the dashed arrow the condensate line, and the wiggly line the effective interaction line including the density fluctuation. 
}
\label{fig2}
\end{figure}

Bimodality can be clearly shown in the density response function in the low-temperature regime (Fig.~\ref{fig1} (a)). A secondary peak is positioned at the foot of a principal peak. This satellite peak structure is distinct from a mere multiparticle excitation~\cite{Glyde1992PRB,Juge1994}, where the total spectrum in the density response function is considered to be a sum of a single sharp peak and a broad continuum~\cite{Miller1962,Woods1965,Woods1978,Griffin1979,Talbot1984,Talbot1988,Stirling1990,Glyde1992PRB,Glyde1995}. The former is originated from the single-particle excitation, and the latter from the multiparticle excitation arising from the 1PI part~\cite{Glyde1992PRB,Juge1994}. In superfluid helium, the latter cannot be negligible since bound states as well as scattering states between maxons/rotons are enhanced because of their very large density of states~\cite{Ruvalds1970,Zawadowski1972,Griffin1973,Manousakis1986,Fukushima1988,Juge1994}. This effect may emerge also in certain classes of ultracold atomic BECs with the roton, such as a dipolar BEC~\cite{Klawunn2009,Blakie2012} as well as a spin-orbit-coupled BEC~\cite{Ozawa2013,Si-Cong2015}.

\begin{figure}[tb] 
\centering
\includegraphics[width=3.2in]{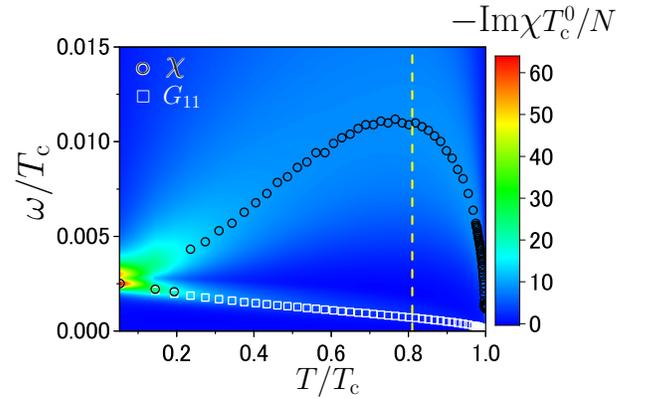}
\caption{
Temperature dependence of density response function. The open-circle represents the maximum peak position of $\chi$, and the open-square that of the single-particle Green's function. Vertical dashed line represents the temperature above which the satellite peak vanishes in the single-particle excitation. The parameters $a n^{1/3} = 10^{-2}$ and $q = 0.01 q_{0}$ are used. 
}
\label{fig3}
\end{figure}

In the present case, the broad 1PI contribution is quite smaller than that of the 1PR part. However, the secondary peak can be clearly seen in the density response function. Since the single-particle excitation is hybridized to the density excitation in the BEC phase, a key to understand the bimodality in the density response function is rather the single-particle excitation. In order to uncover its origin, the single-particle spectral function is studied by selectively including the self-energy contributions $\Sigma_{11(12),{\rm c}}$ as well as $\Sigma_{11(12),{\rm n}}$, where the self-energy $\Sigma_{12,{\rm n}} (p)$ is also considered so as not to eliminate the off-diagonal Green's function in the BEC phase in the case of $\Sigma_{12,{\rm c}}$ eliminated.

The bimodality in the single-particle excitation is absent only in the case where the condensate contributions $\Sigma_{11(12),{\rm c}}$ are not included (Fig.~\ref{fig2}). It indicates that the bimodality is provided by the condensate contributions $\Sigma_{11(12),\rm c}$, which include the interaction effect between the condensate and the quasi-particle in the many-body medium with the multiparticle excitation. A quasi-particle--quasi-particle interaction in the many-body medium is not essential to the bimodality, even if the effective interaction includes the effect of the density-fluctuated medium (line C in Fig.~\ref{fig2}). The effect of the many-body medium is smeared out by quasi-particles with various momenta.

Since the Bogoliubov mean-field approximation can no longer describe the bimodality, which only results in a single sharp peak, the bimodality is an evidence of many-body effect in the BEC, {\it i.e.,} beyond Bogoliubov mean-field effect. 
The satellite peak originated from this many-body effect in the single-particle excitation survives up to $T = 0.81T_{\rm c}$ (Fig.~\ref{fig3}), which is accessible in experiments. 
This satellite peak also survives up to the momentum $q = 0.51q_{0}$ at $T = 0.1T_{\rm c}^{0}$, which corresponds to the wavelength $\lambda = 3.8\mu$m evaluated with experimental values for a homogeneous BEC~\cite{Lopes2017}. 
To briefly conclude, although the mere multiparticle excitation given by the 1PI part does not provide any dominant contribution directly to the total density response function at extremely low temperatures,  this weak multiparticle excitation effect is no longer hidden, but emerges in the density excitation through the single-particle excitation. 

\begin{figure}[tb] 
\centering
\includegraphics[width=3.5in]{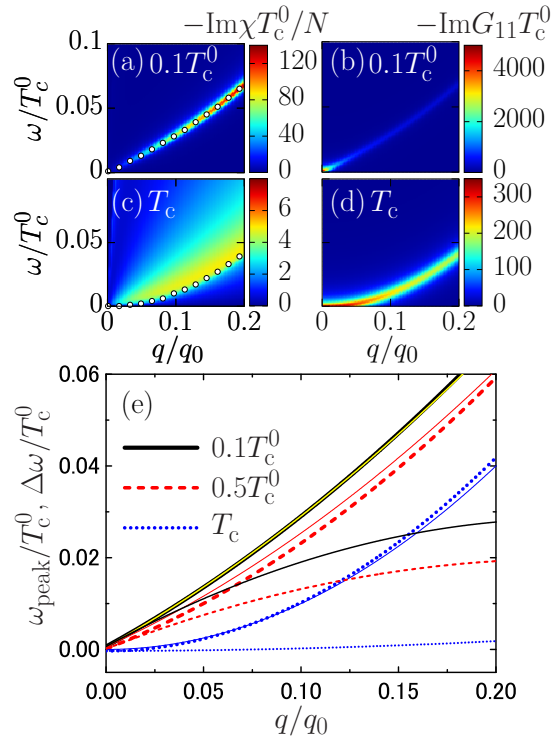}
\caption{
Density response function (a) and single-particle spectral function (b) in the $\omega$-$q$ plane at $0.1T_{\rm c}^{0}$. 
Those at $T_{\rm c}$ [(c) and (d)]. 
Dots in panels (a) and (c) show the peak position of the single-particle spectral function. 
(e) Peak position of the single-particle spectral function $\omega_{\rm peak}$ and their line shift $\Delta \omega$. 
Thick (black solid, red dashed, and blue dotted) lines represent peaks of the single-particle spectral function. Thin lines of them show the corresponding line shift $\Delta \omega$. 
Thin solid lines represent the Bogoliubov excitation spectrum at $0.1T_{\rm c}^{0}$ (yellow),  $0.5T_{\rm c}^{0}$ (red), and  $T_{\rm c}^{}$ (blue). 
We used $a n^{1/3} = 10^{-2}$. 
}
\label{fig4}
\end{figure}

In the BEC at $T=0$, the single-particle Green's function is known to have the same pole as the density correlation function in the low energy regime~\cite{Gavoret1964}. In the low-energy and low-momentum limits, indeed, the density response function is exactly given by~\cite{Gavoret1964} 
\begin{align}
      \chi_{}^{} (p) \simeq & 
      \frac{n}{m} \frac{{\bf p}^{2}}{ \omega^{2} - c^{2} {\bf p}^{2} }, 
\end{align} 
and the leading order of the single-particle Green's function is exactly given by~\cite{Gavoret1964} 
\begin{align} 
      G (p) \simeq & \frac{n_{0} m c^{2}}{n} \frac{ 1 }{ \omega_{}^{2} - c^{2} {\bf p}^{2} } \begin{pmatrix} 1 & -1 \\ -1 & 1 \end{pmatrix} 
\label{eq70}. 
\end{align} 
The sound speed $c$ of the quasi-particle is equal to that given by the thermodynamic compressibility $n/(mc^{2}) = \left ( d n / d\mu \right )_{T}$~\cite{Gavoret1964}. This statement is known to be true for the absolute zero-temperature case, and there has been debate on a nonzero temperature case. We here show that only at extremely low temperatures, peaks of the single-particle Green's function as well as the density response function are identical in numerical resolution; however, those become estranged at higher temperatures (Figs.~\ref{fig3} and~\ref{fig4}). The density response function is given by the sum of both 1PI and 1PR parts $\chi = \chi^{\rm 1PI} + \chi^{\rm 1PR} $, and the 1PR part takes the pole over the single-particle Green's function. However, at higher temperatures, the weight given by the density vertices $\Upsilon$ as well as $\Upsilon^{\dag}$ gets so small that the peak of the density correlation function quantitatively shifts from that of the 1PR part to that of the 1PI part growing relatively. 

We also study the main-peak position $\omega_{\rm peak}$ of the single particle spectral function and its line shift that is defined by $\Delta \omega \equiv \omega_{\rm peak} - {\bf p}^{2}/(2m)$~\cite{Papp2008,Hofmann2017} (Fig.~\ref{fig4}). 
In the low-temperature case ($0.1T_{\rm c}^{0}$), although the many-body effect emerges as the bimodal shape of the single-particle spectral function, its main peak position is well described by the Bogoliubov excitation, even if the many-body effect is included into the self-energy. At higher temperature (0.5$T_{\rm c}^{0}$ and $T_{\rm c}$), the discrepancy between them can be shown, where at the critical temperature in particular, the effective mass of the quasi-particle becomes slightly heavier with the increase of the momentum. 

Notify that the diagrammatic contribution employed in this paper satisfies exact identities. Since the lowest contribution of the regular part shows the infrared divergence $\chi_{\rm R}^{0} (p) \propto - T/ |{\bf p}|$ at nonzero temperatures~\cite{Zwerger2004,Dupuis2011,Watabe2014}, the off-diagonal self-energy satisfies the Nepomnyashchii-Nepomnyashchii identity $\Sigma_{12} (0) = \Sigma_{12,{\rm c}} (0) = 0$~\cite{Nepomnyashchii1978}, which leads the infrared divergence of the longitudinal susceptibility caused by the higher order phase-phase correlation~\cite{Nepomnyashchii1983,Podolsky2011,Watabe2014,Zwerger2004,Dupuis2011,Dupuis2011L}. Because of the same reason of the infrared divergence, the density vertices also satisfy the exact identity $\Upsilon (0) = \Upsilon^{\dag} (0) = 0$~\cite{Nepomnyashchii1978}. This identity leads the exact relation $\lim\limits_{{\bf p} \rightarrow {\bf 0}} \chi_{00}^{\rm 1PR} (0, {\bf p}) = 0$, which indicates that the compressibility zero-frequency sum rule is exactly exhausted by the 1PI part alone~\cite{Gavoret1964,Nepomnyashchii1978}
\begin{align}
      \lim\limits_{{\bf p} \rightarrow {\bf 0}} \chi_{00} (0, {\bf p}) = \lim\limits_{{\bf p} \rightarrow {\bf 0}} \chi_{00}^{\rm 1PI} (0, {\bf p}) = - \frac{n}{mc^{2}}. 
\end{align} 
If the 1PI part were constructed with the lowest order of the regular part $\chi_{\rm R}^{0}$, {\it i.e.,} $\chi^{\rm 1PI} = \chi_{\rm R}^{0}/(1-U\chi_{\rm R}^{0})$, this 1PI part would provide $\lim\limits_{{\bf p} \rightarrow {\bf 0}} \chi_{00}^{\rm 1PI} (0, {\bf p})  = - n / ( mc_{0}^{2})$, because of the infrared divergence of $\chi_{\rm R}^{0}$, where $c_{0} \equiv \sqrt{Un/m}$. This gives an unphysical temperature-independent sound speed $c_{0}$. This problem has been cured in our formalism. By inversely solving the compressibility zero-frequency sum rule, the temperature-dependent sound speed is given by $c = \sqrt{- n / [m \chi^{\rm 1PI} (0)]}$, where $\chi^{\rm 1PI} (0) = - [1 - U \Pi' (0)] / [3 - 2 U \Pi' (0)]$ with 
\begin{align}
      \Pi' (0) \equiv 
       \lim\limits_{{\bf q} \rightarrow {\bf 0}}\Pi' (0,{\bf q}) 
      = 
      \sum\limits_{\bf p} 
      \frac{ \varepsilon_{\bf p}^{2} }{ E_{\bf p}^{2} } 
      \left ( 
      \frac{\partial n_{\bf p}}{ \partial E_{\bf p}} 
      - 
      \frac{1 +2  n_{\bf p}}{2 E_{\bf p}}  
      \right ). 
\label{eq88}
\end{align} 
Here, $n_{\bf p}$ is the Bose distribution function and $E_{\bf p} \equiv \sqrt{\epsilon_{\bf p} (\epsilon_{\bf p} + 2 U n_{0})}$. The stability of the system, represented by $c\in \mathbb{R}$ (as well as $c > 0$), is guaranteed by negativity of $\chi^{\rm 1PI} (0)$ as well as that of $\Pi' (0)$ thanks to the relation $\partial n_{\bf p} / \partial E_{\bf p}  < 0$.

The hidden multiparticle excitation nature unveiled in this paper can be accessible by using spectroscopy for ultracold atoms. The single-particle spectral function can be measured with photoemission spectroscopy~\cite{Stewart2008}; the dynamic structure factor can be measured with Bragg scattering~\cite{Stamper-Kurn1999,Steinhauer2002} as well as a recent ultrahigh finesse optical cavity~\cite{Landig2015}. Ultracold atoms thus offer a prominent platform for studying both the single-particle spectral function and the dynamic structure factor in a BEC. 

This recent advantage of spectroscopy may provide an experimental test of identification of spectra between the single-particle phonon excitation and the density phonon excitation in the BEC. In liquid helium, neutron scattering and light scattering have been successfully uncovered the dynamic structure factor~\cite{Woods1965,Woods1973,Winterling1973,Tarvin1977,Woods1978,Talbot1988,Stirling1990,Diallo2014}. However, since a helium atom is of closed-shell, the single-particle spectral function has not been clearly understood, our understanding of which has been interpretationally extracted from the dynamic structure factor through theoretical knowledge~\cite{griffin1993excitations}. The controllable platform --- the ultracold atoms --- may offer the first direct experimental test of the long-standing problem in the BEC: the identification of spectra between the single-particle excitation and the density excitation, which will deepen our insight of the BEC, and is doubted very recently~\cite{Kita2010,Tsutsumi2016}. 

In summary, multiparticle excitation effects have been investigated on the single-particle spectral function as well as the density response function in a weakly interacting Bose--Einstein condensate (BEC) at nonzero temperatures, with construction of vertex functions satisfying exact identities. Typically below $80\%$ of the critical temperature $T_{\rm c}$, these spectral functions manifest bimodality even without maxons and rotons that have been considered to provide significant contributions to multiparticle excitations. Although the $s$-wave interacting BEC provides a weak multiparticle excitation at extremely low temperatures, this multiparticle excitation effect may not remain hidden. The bimodality in the single-particle excitation is an evidence of the condensate--quasi-particle interaction effect in the density-fluctuated medium with the multiparticle excitation. Visible at extremely low temperatures is identification of spectra between the single-particle and density excitations, where the maximum peak position at 10\% of $T_{\rm c}$ traces the mean-field Bogoliubov spectrum, although the many-body effect can be seen as the satellite peak. This identification between the single-particle excitation and the collective density excitation is the unique nature of the BEC
At higher temperatures, however, it may be invisible since the hybridization becomes weaker between both excitations. 
The results unveiled in this paper will be uniquely accessible by using spectroscopy, such as photoemission spectroscopy, Bragg scattering and ultrahigh finesse optical cavity, in ultracold atoms that will bring our deeper understanding of the single-particle excitation and the collective density excitation in a BEC. 

\begin{acknowledgements}
The author has been supported by JSPS KAKENHI Grant No. (249416, JP16K17774) and Research Fund of Tokyo University of Science. 
\end{acknowledgements}

\appendix*
\section{} 

In this appendix, theoretical details used in this paper, including the self-energy contributions as well as the other building blocks, are summarized. Those are constructed by the Green's function in the Hatree-Fock-Bogoliubov-Popov approximation, given by 
\begin{align}
g^{-1}(p) = i\omega_{n} \sigma_3 - \epsilon_{\bf p} + \mu_{0} - U n_{0} \sigma_{1}, 
\end{align} 
with $\mu_{0} = U n_{0}$ and the Pauli matrices $\sigma_{1,2,3}$. 
The effective interaction $U_{\rm eff}^{0} (p)$ in the density-fluctuated medium, which is used to calculate the self-energy contribution, is described by 
\begin{align}
U_{\rm eff}^{0} (p) \equiv \frac{U}{1 - U \chi_{\rm R}^{0} (p)}, 
\end{align} 
where $\chi_{\rm R}^{0} (p) \equiv \langle f_{0} | \Pi (p) | f_{0} \rangle /2$ is the lowest contribution of the regular part of the density response function. 
Here, 
\begin{align}
\Pi (p) \equiv - T \sum_{q} g (p+q) \otimes g(q)
\end{align} 
is the generalized polarization function, and we have used $\langle f_{0} | \equiv (0, 1, 1, 0)$ and $| f_{0} \rangle \equiv (0, 1, 1, 0)^{\rm T}$. 

The one-particle irreducible (1PI) part is employed as $\chi^{\rm 1PI} (p) \equiv  \chi_{\rm R} (p) / [1 -U \chi_{\rm R} (p)]$, where its regular part is 
\begin{align}
      \chi_{\rm R} (p) \equiv \frac{1}{2} \{  \langle f_{0} | [ \Pi (p) + \Pi (p) \Gamma (p) \Pi (p) | f_{0} \rangle \}, 
      \label{chiR}
\end{align}
with the generalized $T$-matrix $\Gamma (p) \equiv U / [1 - U \Pi (p)]$. 
The BEC-specific density vertices $\Upsilon$ and $\Upsilon^{\dag}$, which includes the single-particle Green's function $G^{} (p)$ into the density response function, are given by 
\begin{align} 
      \Upsilon (p) = & \sqrt{-1} [ G_{1/2} + {\mathcal G}_{1/2}^{\dag} \hat T \boldsymbol{\mathit \gamma}^{} (p)] A (p) , 
      \label{Upsilon}
      \\
      \Upsilon^{\dag} (p) = & \sqrt{-1} [ G_{1/2}^{\dag} 
      + \boldsymbol{\mathit \gamma}^{\dag} (p) \hat T {\mathcal G}_{1/2} ]A(p) , 
\label{Upsilondag}
\end{align}
where three point vertices $\boldsymbol{\mathit \gamma} (q)$ and $\boldsymbol{\mathit \gamma}^{\dag} (q)$ are given by 
\begin{align} 
\boldsymbol{\mathit \gamma} (q)  \equiv { \Gamma} (q) { \Pi} (q) |f_0 \rangle, 
\quad 
\boldsymbol{\mathit \gamma}^{\dag} (q) \equiv \langle f_0 | { \Pi} (q) { \Gamma} (q), 
\end{align}
the condensate Green's functions $G_{1/2} \equiv \sqrt{- n_{0}} (1, 1)^{\rm T}$, $G_{1/2}^{\dag} \equiv \sqrt{- n_{0}} (1, 1)$, ${\mathcal G}_{1/2} \equiv \sqrt{- n_{0}} \eta_{g}$, and ${\mathcal G}_{1/2}^{\dag} \equiv \sqrt{- n_{0}} \hat \eta_{g}^{\rm T}$, as well as a vertex coefficient $A(p) \equiv 1 / [1 - U \chi_{\rm R}^{0} (p)]$. We have introduced
\begin{align} 
      \hat T = 
      \begin{pmatrix}
            1 & 0 & 0 & 0 \\ 0 & 0& 1 & 0 \\ 0 & 1 & 0 & 0 \\ 0 & 0& 0 & 1
      \end{pmatrix}, 
            \quad
            \hat \eta_{\rm g} = 
      \begin{pmatrix}
            1 & 0  \\ 1 & 0 \\ 0 & 1 \\ 0 & 1
      \end{pmatrix}, 
\end{align}
and $\hat \eta_{g}^{\rm T}$ is the transpose of $\hat \eta_{g}$. 

In this paper, the self-energy contributions are considered as the Hatree-Fock-Bogoliubov-Popov-type diagram, 
which are given by $\Sigma_{11} (p) = \Sigma_{11, \rm c} (p) + \Sigma_{11,\rm n} (p)$ as well as $\Sigma_{12} (p) = \Sigma_{12,\rm c} (p)$, 
where 
\begin{align} 
\Sigma_{11, \rm c} (p) =  & U_{\rm eff}^{0} (0) n_{0}  + U_{\rm eff}^{0} (p) n_{0}, 
\\ 
\Sigma_{11, \rm n} (p) = & U_{\rm eff}^{0} \tilde n -T \sum_{q} U_{\rm eff}^{0} (q) g_{11} (p + q), 
\\ 
\Sigma_{12, \rm c} (p) = & U_{\rm eff}^{0} (p) n_{0}, 
\end{align}
with $\tilde n \equiv - T \sum_{p} g_{11} (p) e^{i \omega_{n} \delta}$. 
We also considered the self-energy contributions as the Hatree-Fock-Bogoliubov-type diagram in Fig.~\ref{fig2}, where $\Sigma_{12} (p)$ is replaced by $\Sigma_{12} (p) = \Sigma_{12, \rm c} (p) + \Sigma_{12,\rm n} (p)$ with 
\begin{align} 
\Sigma_{12,{\rm n}} (p) = - T \sum_{p} U_{\rm eff}^{0} (p) g_{12} (p +q). 
\end{align}

\bibliographystyle{apsrev4-1}
\bibliography{library.bib}

\end{document}